\documentclass[preprint,preprintnumbers,amsmath,amssymb]{revtex4}

\usepackage{graphicx}
\usepackage{bm}

\begin{document}

\preprint{CAS-KITPC/ITP-XXX}

\title{~\\ \vspace{2cm}
Non-Commutativity, Teleology and GRB Time Delay
\vspace{1cm}}

\author{Miao Li}\email{mli@itp.ac.cn}
\author{Yi Pang}\email{yipang@itp.ac.cn}
\author{Yi Wang}\email{wangyi@itp.ac.cn}
\affiliation{Kavli Institute for Theoretical Physics China,
Key Laboratory of Frontiers in Theoretical Physics,
Institute of Theoretical Physics, Chinese Academy of Sciences,
Beijing 100190, P.R.China\vspace{2cm}
}%

\begin{abstract}
We propose a model in which an energy-dependent time delay of a
photon originates from space-time non-commutativity, the time delay
is due to a  noncommutative coupling between dilaton and photon. We
predict that in our model, high energy photons with different
momentum can either be delayed or superluminal, this may be related to
a possible time delay reported by the Fermi LAT and Fermi GBM Collaborations.
\end{abstract}

\maketitle

\section{Introduction}

Recently, the Fermi LAT and Fermi GBM Collaborations reported a time
delay effect for high energy photons \cite{GRB080916C}. The time delay effect was
from GRB 080916C, which is an extremely energetic gamma-ray burst located at $z=4.35\pm 0.15$.
It is reported that the $13.22^{+0.70}_{-0.54}$ GeV signal was delayed for 16.54 seconds.
Similar time delay results for high energy photons were also reported by MAGIC
\cite{MAGIC} and HESS \cite{HESS} previously.

Several interpretations have been proposed for the time delay. For example, the time delay
may originate from some astrophysics mechanisms, such as two different pairs of colliding
shells, or high energy photons were attenuated until the emitting region became optically thin.

Alternatively, the time delay may be due to Lorentz violation in fundamental physics. Since the time the photon takes
to travel from the GRB to us is comparable with the age of the universe, even very tiny modification of
standard physics may show up in the time delay. For example, if one assumes
that the delay effect is from a modified speed of light \cite{Ellis:2005wr}
$
  c(E)\simeq c_0 (1-E/M_{\rm QG})~,
$ then the bound for quantum gravity energy scale $M_{\rm QG}$
becomes $M_{\rm QG}>1.3\times 10^{18}$GeV. As another example, in
\cite{Li:2009tt}, it is reported that the delay effect can come from
string theory in D-particle backgrounds.

In this letter, we investigate the time delay effect from space-time
non-commutativity \cite{SSTY}. We find that space-time
non-commutativity from oscillating dilaton-photon coupling can
produce either larger or smaller speed of light, depending on the
momentum. We also point out that the oscillating dilaton-photon
coupling may also be measured in future collider experiments. Finally, we must mention
that the Lorentz violating effect of space-time noncommutativity can not
be mimicked by a finite Lorentz violating operator.

\section{Non-Commutative Dilaton-Photon Coupling}

In \cite{ssur}, the authors point out that the physical time and space
coordinates for strings and D-branes should satisfy the following uncertainty relation:
\begin{equation}
  \Delta t_p \Delta x_p \geq l_N^2 ~.
\end{equation}
This space-time uncertainty relation can be realized with non-commutative space-time with
non-commutativity length scale $l_N$, by replacing the usual product between fields with the
*-product. In \cite{Brandenberger:2002nq}, this space-time
non-commutativity is applied to cosmology. Subsequent developments and applications of this
cosmic non-commutativity can be found in \cite{nc} and references theirin.

To realize space-time non-commutativity in cosmology background, one
needs to introduce time coordinate $\tau$, with metric
\begin{equation}
  ds^2=-a^{-2}d\tau^2+a^2 d{\bf x}^2~,
\end{equation}
so that the uncertainty relation in terms of $\tau$ and $x$ is still time independent.
To investigate time delay, we propose the following action for dilation-photon coupling:
\begin{equation}\label{actiondilaton}
  S=-\frac{1}{4}\int d^3 x d\tau \sqrt{-g}g^{\alpha\mu}g^{\beta\nu}F_{\alpha\beta}*\Phi*F_{\mu\nu}~,
\end{equation}
where $\Phi=\Phi(\tau)$ is the time varying dilaton field, relating
to the fine structure constant as $\Phi \sim 1/\alpha$. The
*-product is defined as
\begin{equation}
  (f*g) (x,\tau)=\left.
e^{-i l_N^2 (\partial_{x}\partial{\tau'}-
\partial_{\tau}\partial_{y})}f(x,\tau)g(y,\tau')\right|_{y=x, \tau'=\tau}~,
\end{equation}
where we shall identify $x$ and $y$ with the radial coordinate $r$. There is no way to define
an explicit *-product without breaking translational invariance, a trick to replace *-product
in order to preserve translational invariance is proposed in \cite{Brandenberger:2002nq}, we assume
that a similar trick is applicable to a vector field.
In the action \eqref{actiondilaton}, we do not consider the non-commutative product between
the scale factor and the gauge field. It is because, as we will show in the next section,
the effect of non-commutativity from scale factor is too small to detect.

We work in the Coulomb gauge $A_0=0$, $\partial^i A_i=0$. In this gauge, the real degree of freedom of photon
becomes $A^r$ ($r=1, 2$), with
\begin{equation}
  A_i({\bf x})=\int\frac{d^3 k}{(2\pi)^3}\sum_{r=1,2}\epsilon_i^r({\bf k})A^r_{\bf k} e^{i {\bf k}\cdot {\bf x}}~.
\end{equation}
After applying the *-product, the action \eqref{actiondilaton} becomes
\begin{equation}
  S=\frac{1}{2}\int d\tau \frac{d^3 k}{(2\pi)^3}\sum_{r=1,2}
\left\{
a^2 \tilde\Phi \partial_\tau A_{\bf k}^r \partial_\tau A_{- \bf k}^r
- k^2 a^{-2} \tilde\Phi A_{\bf k}^r A_{- \bf k}^r
\right\}~,
\end{equation}
where
\begin{equation}
  \tilde\Phi=\frac{1}{2}\left[ \Phi(\tau+l_N^2k)+\Phi(\tau-l_N^2k) \right]~.
\end{equation}
The equation of motion of $A_{\bf k}^r$ can be written as
\begin{equation}\label{eomtau}
  \partial_\tau \left(a^2\tilde\Phi\partial_\tau A_{\bf k}^r\right) +k^2 a^{-2} \tilde\Phi A_{\bf k}^r=0~.
\end{equation}
To solve above equation conveniently, we define another time
coordinate $\rho$ as $\partial_\rho=a^2\tilde\Phi\partial_\tau$,
then Eq. \eqref{eomtau} becomes
\begin{equation}\label{eomrho}
  \partial_\rho^2 A_{\bf k}^r+k^2\tilde\Phi^2 A_{\bf k}^r=0~.
\end{equation}
The solution of this equation can result in a modified group velocity for photons. To
proceed, we propose an explicit time dependence of dilaton
\begin{equation}
  \Phi=\Phi_0+\Phi_1 \cos(\omega \tau)~.
\end{equation}
We assume that the dilaton oscillates very fast, because this can produce
large time dependence without significantly changing the fine
structure constant in low energy experiments, this implies that
$\Phi_1\ll \Phi_0$. With these assumptions, we have
\begin{equation}
  \tilde\Phi=\Phi_0+\Phi_1 \cos(\omega\tau)\cos(\omega l_N^2 k)~.
\end{equation}
When $k$ is very small compared with the frequency $\omega$, the $\cos (\omega\tau)$
term can be averaged over, so does not make any difference from that without oscillation.

When $\Phi_0 k \gg a^2 \Phi_1 \omega$, Eq. \eqref{eomrho} can be solved order by order using the WKB approximation as
\begin{align}
  A_{\bf k}^r &= \varphi_{\bf k}^r \exp
\left\{i\int dt \frac{k}{a} \left[
1+\frac{1}{8}\frac{a^4\Phi_1^2\omega^2}{\Phi_0^2k^2}\left(
\sin^2(\omega\tau)-2\cos^2(\omega\tau)\right)\cos^2(\omega l_N^2
k)\right] \right\} \nonumber\\& \equiv \varphi_{\bf
k}^r\exp\left\{i\int \Omega(k,t)dt \right\}~,
\end{align}
where we have neglected the oscillating terms which become zero after the time integration.

\begin{figure}
  \center
  \includegraphics[width=0.6\textwidth]{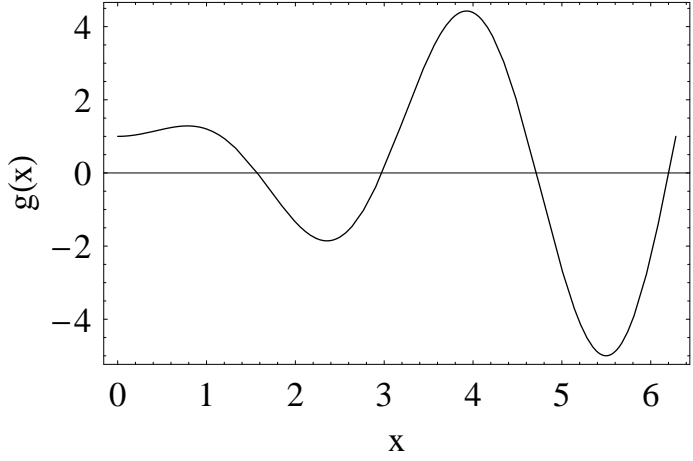}
    \caption{\label{gx} This figure plots g(x) as a function of x. Note that the sign of $g$ can change.}
\end{figure}

The group velocity takes the form
\begin{equation}
  v_g(k)=\frac{d\Omega}{dk}=\frac{1}{a}\left\{1+\frac{1}{8}
\frac{a^4\Phi_1^2\omega^2}{\Phi_0^2k^2}\left(2\cos^2(\omega\tau)-\sin^2(\omega\tau)\right)
g(\omega l_N^2 k)
\right\}~.
\end{equation}
where
\begin{equation}
  g(\omega l_N^2 k)\equiv\left(\cos^2(\omega l_N^2 k)+ \omega l_N^2 k \sin(2\omega l_N^2 k)\right)
\end{equation}
In Fig. \ref{gx}, we plot $g(\omega l_N^2 k)$ as a function of
$\omega l_N^2 k$. One can see that the sign of $g$  changes as a function. So
a group velocity larger than or smaller than speed of light can both
be achieved with non-commutativity.

The comoving distance from the GRB to us can be calculated as
\begin{equation}
  \Delta x = \int_{t_{\rm GRB}}^{t_0} \frac{dt}{a} = \int_{t_{\rm GRB}}^{t_k}
  v_g(k)dt~,
\end{equation}
where $t_k$ is the arriving time of the photon with momentum $k$.

Then the time delay can be written as
\begin{equation}
  \Delta t \equiv t_k-t_0 =\frac{1}{v_g(k)}\int_{t_{\rm GRB}}^{t_0} dt \left[
\frac{1}{a}-v_g(k)
\right]~.
\end{equation}
In the case under consideration, $\omega^{-1}$ is much smaller than cosmic
time scale. So one can average over $\cos(\omega\tau)^2$ and
$\sin^2(\omega\tau)$ in the integrand first, in other words, to
replace them by their average value $\frac{1}{2}$. Finishing the
time integration, one gets
\begin{equation}
  \Delta t= -\frac{g(\omega l_N^2 k)}{48 H_0 \sqrt{\Omega_{\Lambda0}}} \frac{\Phi_1^2 \omega^2}{\Phi_0^2 k^2}
 \left[ f(1)-f\left(\frac{1}{(1+z_{\rm GRB})^3}\right)\right]~,
\end{equation}
where today's scale factor has been chosen to be 1, and the  $f(x)$
is defined as
\begin{equation}
 f(x)\equiv \int \frac{dx}{\sqrt{1+\frac{\Omega_{m 0}}{x\Omega_{\Lambda0}}}}
 =\sqrt{x\left(\frac{\Omega_{m 0}}{\Omega_{\Lambda
0}}+x\right)}-\frac{\Omega_{m 0}}{\Omega_{\Lambda 0}}  \log \left(
2\sqrt{x}+2\sqrt{\frac{\Omega_{m 0}}{\Omega_{\Lambda 0}}+x} \right)~.
\end{equation}
Inserting $z_{\rm GRB}=4.35$, $\Omega_{\Lambda0}=0.721$ and $\Omega_{m0}=0.279$, one has
\begin{equation}
  f(1)-f\left(\frac{1}{(1+z_{\rm GRB})^3}\right) = 0.692~.
\end{equation}
Further applying $H_0^{-1}=4.40\times 10^{17}$s, we have
\begin{equation}
  \Delta t = -7.47\times 10^{15} g(\omega l_N^2 k)\frac{\Phi_1^2\omega^2}{\Phi_0^2 k^2} ~{\rm s}=16.54 {\rm s}~.
\end{equation}
Given $k=13.22$GeV, we still have two free parameters to fix in
our model: the combination $\omega l_N^2 k$ and the ratio
$\Phi_1/\Phi_0$. For example, taking $\omega l_N^2 k=2$, and
$\Phi_1/\Phi_0=10^{-10}$, we have
\begin{equation}
  \omega=5.37{\rm TeV}~,\qquad l_N^{-1}=188{\rm GeV}~.
\end{equation}
One can check that to fit the momentum and time delay, the WKB approximation
is always very robust.

Note that $\Delta t$ is a oscillating function of $\omega l_N^2 k$.
Then our model predicts that for high energy photons with different
momenta, some will reach us earlier than the low energy photons,
while some will delay. The time $-\Delta t$ as a function of $\omega
l_N^2 k$ should be proportional to $g(\omega l_N^2 k)$, as shown in
Fig. \ref{gx}.

It is also remarkable that under the commutative limit
$l_N\rightarrow0$, the remaining oscillating effect still
contributes to the time delay. Specifically, in the restriction of
WKB approximation, photons with larger momenta will arrive later
than those with smaller momenta. However, they are all superluminal
compared with the photon carrying too small momentum, then
unaffected by the highly oscillating dilaton.

Finally, we would like to discuss the bound for the above parameters, and whether they are detectable
on colliders. If we want the non-commutative scale be
 universal, in other words, there are some other
non-commutative effects around the non-commutative scale, then $l_N^{-1}$ should be greater than the
well examined energy scale on colliders, say, 100GeV. In this case, $\Phi_1/\Phi_0$ is always too
small to be measurable on colliders.

On the other hand, if the non-commutative scale applies only for the photon-dilaton coupling, then
$l_N^{-1}$ can be much lower than 100GeV, and one can expect some oscillating behavior on the fine structure
constant measured on linear colliders.

\section{Non-Commutativity from the Metric}

In this section, we consider the effect of non-commutativity from the graviton-photon coupling. We find that the
contribution to the time delay is suppressed by powers of the Hubble parameter. So this effect is too small to be observable.
This verifies the assumption in the previous section that we only need to consider the non-commutativity
from dilaton-photon coupling.

The action for non-commutative graviton-photon coupling is a direct generalization of
\cite{Brandenberger:2002nq}:
\begin{equation}
  S=-\frac{1}{4}\int d^3x d\tau F_{\alpha\beta}*\sqrt{-g} g^{\alpha\mu}g^{\beta\nu}*F_{\mu\nu}~.
\end{equation}
After mode expansion and applying the *-product, we have
\begin{equation}
  S=\frac{1}{2}\int d\tau \frac{d^3k}{(2\pi)^3}\sum_{r=1,2}
\left\{
\beta^+ \partial_\tau A_{\bf k}^r \partial_\tau A_{-\bf k}^r-k^2\beta^- A_{\bf k}^rA_{-\bf k}^r
\right\}~,
\end{equation}
where
\begin{equation}
  \beta^{\pm}=\frac{1}{2}\left[ a^{\pm 2}(\tau+l_N^2k)+a^{\pm 2}(\tau-l_N^2k)  \right]~.
\end{equation}
The equation of motion of $A_{\bf k}^r$ takes the form
\begin{equation}
  \partial_\eta^2 A_{\bf k}^r +k^2 \beta^+\beta^- A_{\bf k}^r=0~,
\end{equation}
where $\eta$ is defined as $\partial_\eta=\beta^+\partial_\tau$.

To expand up to the leading order of $l_N$ and to the leading order
of WKB approximation, one obtains the solution
\begin{equation}
  A_{\bf k}^r = \varphi_{\bf k}^r \exp \left\{
i k  \int d\eta \left(1+ \frac{2l_N^4 k^2 H^2}{a^2}\right)
\right\}~.
\end{equation}
One finds that qualitatively the correction term is
double-suppressed by the scale of non-commutativity and the Hubble
parameter, thus this change of dispersion relation can not be
observable in GRB experiments. But quantitatively this
non-commutative graviton-photon coupling always gives a superluminal
result.

One could assume there is small oscillation in the scale factor, due
to some background field oscillating around its potential. Then larger
effects of non-commutativity may come out. We shall not investigate this
possibility in detail in this paper.

\section{Conclusion}

To conclude, we investigated the non-commutative oscillating dilaton-photon coupling
and its effect on high energy photons. We find that both time delay and superluminal
propagation of high energy photons can be achieved. This can explain the recent
time delay effect from GRB 080916C.

We also investigated the effect of non-commutativity from scale factor. We find that
the non-commutativity from scale factor is too small to be detectable.

\section*{Acknowledgments}
One of the authors Yi Pang would like to thank Zheng Yin for e-mail
communication. We thank R. X. Miao for pointing  out a typo for us.

\end{document}